# Development of a Window Based Security System for Electronic Data Interchange

[1]Achimugu Philip [2]Oluwagbemi Oluwatolani and [3]Abah Joshua

**Abstract-**The Electronic Data Interchange (EDI) is the exchange of standardized documents between computer systems for business use. The objective of this study is to make Electronic Data Interchange secure to use and to eliminate human intervention in the transfer of data between business partners so that productivity and efficiency can be improved and also promote its usage between two or more trading organizations. This paper provides an overview of EDI by describing the traditional problems of exchanging information in business environments and how the EDI solves those problems and gives benefits to the company that makes use of EDI. This paper also introduces the common EDI Standards and explains how it works, how it is used over the internet and the security measures implemented. The system was executed on both local area network and wide area network after a critical study of the existing EDI methods and also implemented using VB.Net programming language. Finally, an interactive program was developed that handles the transfer of files, with special attention to the security of the items that are being transferred from one computer workstation to another.

**Index Terms-**EDI, Information, Standards, Security, Business

———————————— ◆ ————————————

## 1 Introduction

Traditional electronic data interchange (EDI) has been evolving for approximately 25 years and has truly become the paperless environment that is so often talked about. EDI is a complicated mixture of three disciplines: business, data processing, and data communications [10].

Since EDI is commonly defined as the direct computer-to-computer exchange of standard business forms, it clearly requires a business process. Because the key idea involved is the exchange of documents that allow a business application to take place without human intervention, data processing is clearly necessary for application processing. Data communication is then necessary for the exchange to take place. It is the marrying of these three disciplines that allows the "paperless trading" that comprises EDI technologies. Besides the three career disciplines that are internal to the organization, three other issues are important for EDI trading to take place: standardization of formats, security, and value-added networks (VANs).

Electronic data interchange can be used to transmit documents electronically such as invoices, purchase orders, receipts, shipping documents, and other standard business correspondence between organizations and business partners. EDI can also be use to transmit financial information and payment in electronic funds transfer (EFT). Because of these, the functions of EDI becomes more and more significant nowadays especially with the growth of electronic commerce over the world. It is important for us to understand how the EDI works and improves the traditional way of exchanging information between trading partners, so that the productivity and efficiency can be increased.

This paper describes the basic problem and solution in the traditional business environments, which is paper-based system to exchange information. It introduces some

————————————————
1. *Achimugu Philip is with the Department of Computer Science of Lead City University, Ibadan, Nigeria.*
2. *Oluwagbemi Oluwatolani is with the Department of Computer Science of Lead City University, Ibadan, Nigeria.*
3. *Abah Joshua is with the Computer Engineering Department of the University of Maiduguri, Maiduguri, Nigeria.*

worldwide EDI standards. The paper also discusses how the EDI works and the security measure implementation.

Electronic Data Interchange (EDI) may be most easily understood as the replacement of paper-based purchase orders with electronic equivalents. It is actually much broader in its impacts are far greater than mere automation. EDI offer the prospect of easy and cheap communication of structured information throughout the corporate community, and is capable of facilitating much closer integration among hitherto remote organizations. Therefore, EDI can be defined as the exchange of documents in standardized electronic form, between organizations, in an automated manner, directly from a computer application in one organization to an application in another [8].

## 2. Preparation of Electronic Documents

The first step in the transaction of EDI is the acquisition of information. The way to acquire the required information is the same as the way to do it in the traditional system. However, instead of printing out the data on paper as in the traditional way, an electronic file or database to store this information is developed. In the case of organizations who already use computer to issue their documents like purchase orders, they may already have some sort of databases that store those information. Therefore, they can start with the steps described below:

### 2.1 Outbound Translation

The next step is to translate the electronic file or database into a standard format according to the specification of the corresponding document. The resulting data file should contain a series of structured transaction related to the purchase order for example. If more than one company is involved in the particular transaction, individual file should be produced for each of them.

### 2.2 Communication

Then the computer should connect and transmit those data file to the pre-arranged *Value Added* Network (VAN) automatically. The VAN should then process each file and route to the appropriate electronic mailboxes according to the destination set in the file.

### 2.3 Inbound Translation

The designated organizations should be able receive the file in their respective electronic mailboxes constantly and then reverse the process by translating the file from the standard format into the specific format required by the organization's application software.

### 2.4 Processing the Electronic Documents

The internal application system of the designated organization can process the received documents now. All the resulted documents corresponding to the received transaction would use the same process or steps to transmit back to the transaction initiator. The whole cycle of the electronic data interchange can then be completed.

One of the technological fields required to implement EDI is data processing. Data processing allows the EDI operation to take information that is resident in a user application and transform that data into a format that is recognizable to all other user applications that have an interest in using the data. In the EDI environment, data processing will handle both outgoing and incoming data, as depicted in Figure 1.

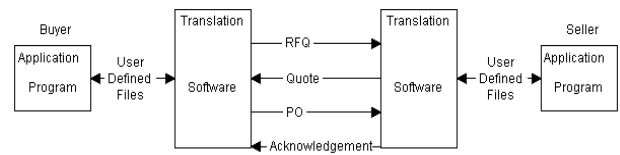

**Figure 1:** Data Processing and EDI

**Source:** [11]

### 2.5 Value Added Networks

Large communications networks have been set up or developed to provide post box (or electronic mail box) services between EDI users. These networks are known as Value Added Networks (VANs) or clearing centers and are built and run by the public telephone companies or large computer suppliers.

An organization who wishes to send EDI message dials up the VAN and deposits packets of which includes the electronic address of the recipient. The VAN takes the data from the post box and sorts it into the recipients' mail boxes ready for those users to collect or acknowledge. Thus at the VAN, each EDI user have its own mail box from which it retrieves messages. Their VANs provide other basics; such as, message tracking and the ability to link other types of computers that could not otherwise link directly. By tracking message through the network, VANs record whether and when messages arrive, as well as whether they are transferred or picked up. VANs have developed the capability to exchange data with a wide variety of different computers using the appropriate communication protocols.

As long as any two users can individually link to the chosen VAN, they will be able to send messages to each other through the VAN even though they may not be able to link directly to each other. A Value Added Network provides an easy means of accessing many other users over a computer network and it also correctly sorts the messages it transit like a postal service. There are many different VANs around the world, some extending to so many countries, others providing national coverage and yet others which serve only a particular industry or community. In the early days of EDI, a user was obliged to subscribe to as many different VANs as its partners (usually customers) required. VANs are beginning to subscribe to more than one network.

## 2.6    Direct Connections

Some large organizations operate their computers most hours of most days and so have decided that they will not use the value added networks. Instead they have chosen to link directly to their customers and suppliers. These direct links can use normal telephone lines with modems at each end or special links, which have been designed to carry digital data. The service behaves like an electronic postal service where letters for lots of address can be put into a post box and the mail service will sort them and send them to the right destination.

A packet switching service is more like a postal service by telephone where a single delivery from the postman can contain letters from a variety of different people. An intermediate form of link is called Packet Switching. There is a single link between each company and a packet exchange network provided by the telephone companies. A terminal then sends information for many different destinations via a line all interleaved together. The exchange network sorts out the route for each packet and sends it to the correct location. There is no mailbox in this particular option, which means that the receiving terminal has to be listening and operating at the same time as the sender is transmitting information to it.

## 2.7    Security

One of the major roles that is provided by the data communications technology is the ability to apply security to EDI transactions so that the transactions will not be tampered with or observed, depending on the level of security needed (Canis, 1995). The security modules that are discussed in this section are depicted in Figure 2.

## 2.8    Confidentiality

This requires that all communications between parties are restricted to the parties involved in the transaction. This confidentiality is an essential component in user privacy, as well as in protection of proprietary information and as a deterrent to theft of information services. Confidentiality is concerned with the unauthorized viewing of confidential or proprietary data that one or both of the trading partners does not want known by others. Confidentiality is provided by encryption. Encryption is the scrambling of data so that it indecipherable to anyone except the intended recipient. Encryption prevents snoopers, hackers, and other prying eyes from viewing data that is transmitted over telecommunications channels. There are two basic encryption schemes, private-key and public-key encryption. Encryption, in general, is cumbersome and expensive. However, Private-key encryption requires that both sending and receiving parties have the same private-encryption keys. The sender encrypts the data using his key. The receiver then decrypts the message using his identical key as shown in Figure 3. There are several disadvantages to private-key encryption. In order to remain secure, the keys must be changed periodically and the users must be in synch as to the actual keys being used; while Public-key encryption is gaining wide spread acceptance as the preferred encryption technology. With public-key encryption, a message recipient generates a matched set of keys, one public key and one private key. The recipient broadcasts the public key to all senders or to a public location where the key can be easily retrieved. Any sender who needs to send the receiver an encrypted message uses the recipient's public key to encrypt the message. The private key, which is held in private by the recipient, is the only key that can decipher messages encrypted with the matched public key. This schema requires that the private key cannot be generated from the public key. Public key technology is the direction encryption technology is currently headed. With the advent of X.500, databases will be built to store public keys and enhance the technology significantly.

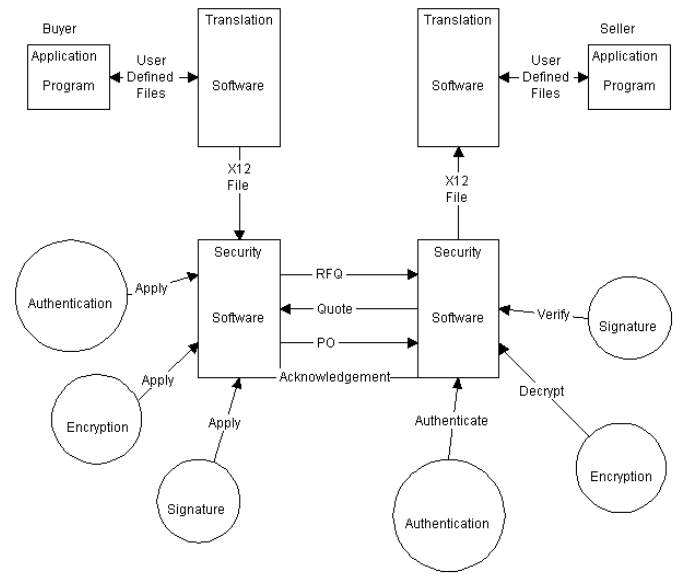

**Figure 2:** Data Communications Security

**Source:** [3]

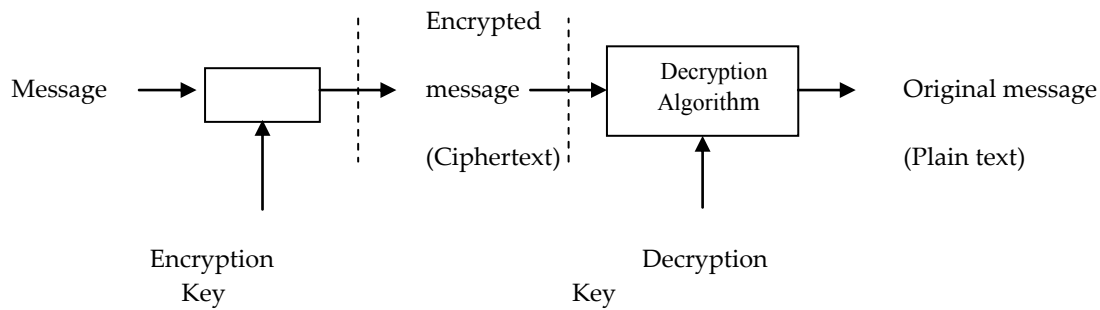

**Figure 3:** Encryption/Decryption Process

## 2.9    Authentication

Both parties should feel comfortable that they are communicating with the party with whom they think they are doing business. A normal means of providing authentication is through the use of passwords. The latest technology to provide authentication is through the use of digital certificates that function much like ID cards. The digital certificate has multiple functions, including browser authentication.

## 2.10    Data Integrity

Data sent as part of a transaction should not be modifiable in transit. Similarly, it should not be possible to modify data in storage. Data integrity is a guarantee that what was sent by the sender is actually what is received by the receiver. This is necessary if there is a need to ensure that the data has not been changed either inadvertently or maliciously. However, authentication schemes do not hide data from prying eyes. Providing data integrity is generally cumbersome and not used unless one of the trading partners requires it. The normal mechanism for acquiring data integrity is for the sender to run an algorithm against the data that is being transmitted and to transmit the result of the algorithm separately from the transmission. Upon receipt of the transmission, the receiver runs the identical algorithm and then compares the results. If the results are identical, then data has not been modified.

## 2.11 Non-repudiation

Neither party should be able to deny having participated in a transaction after the fact. The current technology ensures this through the use of digital signatures. Electronic signatures are the computerized version of the signature function. Signatures are needed in some business applications for authorization purposes. For example, a contracting officer may have a specified spending limit, say $25,000. If that contracting officer decides to place an order for $30,000, the seller may not have the authority to fill the order because the signature of the contracting officer's supervisor is needed on all orders over $25,000. The authorization limits normally will have been agreed upon through a trading partner agreement. A digital signature algorithm can be used to generate digital signatures. The digital signature itself is used to detect unauthorized modification to data and to authenticate the identity of the signature. The digital signature is also useful to the recipient as a non-repudiation device whereby the recipient can prove to a third party that the signature was in fact generated by the signatory. Thus the signatory cannot repudiate the signature at a later date.

## 2.12 Connectivity

VANs establish communications paths between their customers and with other VANs. By using these services a business does not have to worry about the myriad of communications complexities from having trading partners using different hardware, software, and transport mechanisms. The typical buyer-VAN-seller connection is depicted in Figure 4.

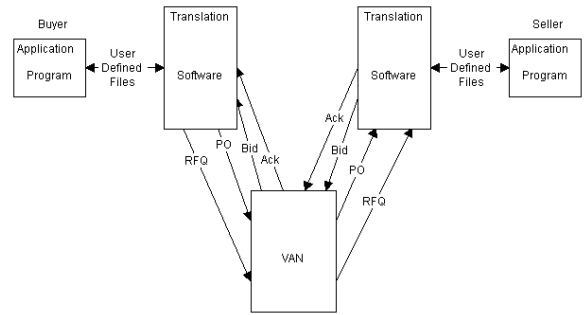

**Figure 4:** Value-Added Network Connection

**Source:** [1]

Likewise, EDI software is not inexpensive. A business with an X12 translator still needs personnel on board that understand X12 and can use the software effectively. Value-added services offer the traditional VAN services and the translation services required to create an X12 file. These services allow the typical business to enter the EDI arena at minimal cost and maximum efficiency.

## 2.13 Delivery

Mailbox software is the most important feature offered by VANs. The electronic mailbox is used for both store-and-retrieve and store-and-forward operations. In both cases, the sender of the EDI message transmits the electronic message to the VAN on its own time schedule. The VAN then acts on the message depending on whether the service is store-and-retrieve or store-and-forward.

Store-and-retrieve service allows the VAN to store the message in the receiver's mail box. The receiver then retrieves messages based on the needs and schedules of the receiver. This service enables the sender and receiver to communicate, but at different times of the day, instead of simultaneously.

Store-and-forward service allows the VAN to forward messages to the receiver when the business need is not for immediate or event-driven notification. Event-driven mailbox services can be handled by forwarding of the message to the receiver or by immediate notification from the VAN to the receiver that a message has been stored that meets the prearranged criteria for event-driven notification.

## 2.14  Value Added Network (VAN) SECURITY

Generally, a VAN provides security at several levels for its mailbox customers. Access control is normally provided by a login and password sequence [1]. Messages are screened for the individual customer to ensure that they were sent by authorized trading partners of the customer. This service also checks for message types and formats, and ensures they are acceptable to the customer. Some VANs offer cryptography services. The cryptography is used to authenticate and encrypt messages to ensure confidentiality. This service requires that the encryption be done at the customer site to be of any real value.

## 2.15  Audit and Control

One of the features a VAN can offer a customer is a usage accounting data option whereby the VAN reports how much traffic comes to the customer in a given time period. Transmission status reports to clarify status of an individual transaction are also available [3]. Many trading partners require acknowledgment for transactions received, and VANs can provide automatic sending of acknowledgments. The VAN can also track the transaction traffic. If specific transactions need to be tracked, the VAN can provide an audit trail of the requested data.

## 2.16 Value-Added Services

In the typical EDI implementation, both sender and receiver employ the services of a VAN because it eliminates the need to support different communications configurations between themselves and their trading partners. Using VANs also reduces the cost of communications equipment and staff to support the multiple configurations. Still, not all trading partners will use the same VANs. This is not an issue because VANs interconnect regularly with each other. The standard VAN interconnection is through bisynchronous modem connections. Most VANs offer translation services so that customers do not have the need to purchase or maintain translation software. Normally, if these services are used, the customer will supply the formats for the data and the VAN will map the data itself. VANs have the capability to respond to presence of data and can fax or e-mail a notification to the customer if data is in the customer's mailbox.

## 2.17  Effects and Level of Automation

The benefits associated with EDI often causes overblown expectations. EDI, in and of itself, is just another way to format and transfer data. The real use of EDI and the amount of value to be gained from its implementation depend upon whether or not EDI is integrated into the overall data processing effort of the organization. Furthermore, the effects of EDI depend greatly on the level of automation within an organization. If the organization is only using EDI to send data in a format required by a trading partner, the effect is much more limited than if EDI is integrated into the back-end processes of the organization. EDI applications that are fed by back-end processes and the databases that support these processes and then, in turn, feed the EDI data received back into the databases and back-end processes have a huge impact on the total level of automation within organization. The well-known list of EDI-related benefits-lower costs, higher productivity, and reduced order-cycle times-is attainable. But if the automation level of the organization is not high and is not integrated, the effects of EDI will be lessened considerably.  Figure 5 demonstrate how EDI works.

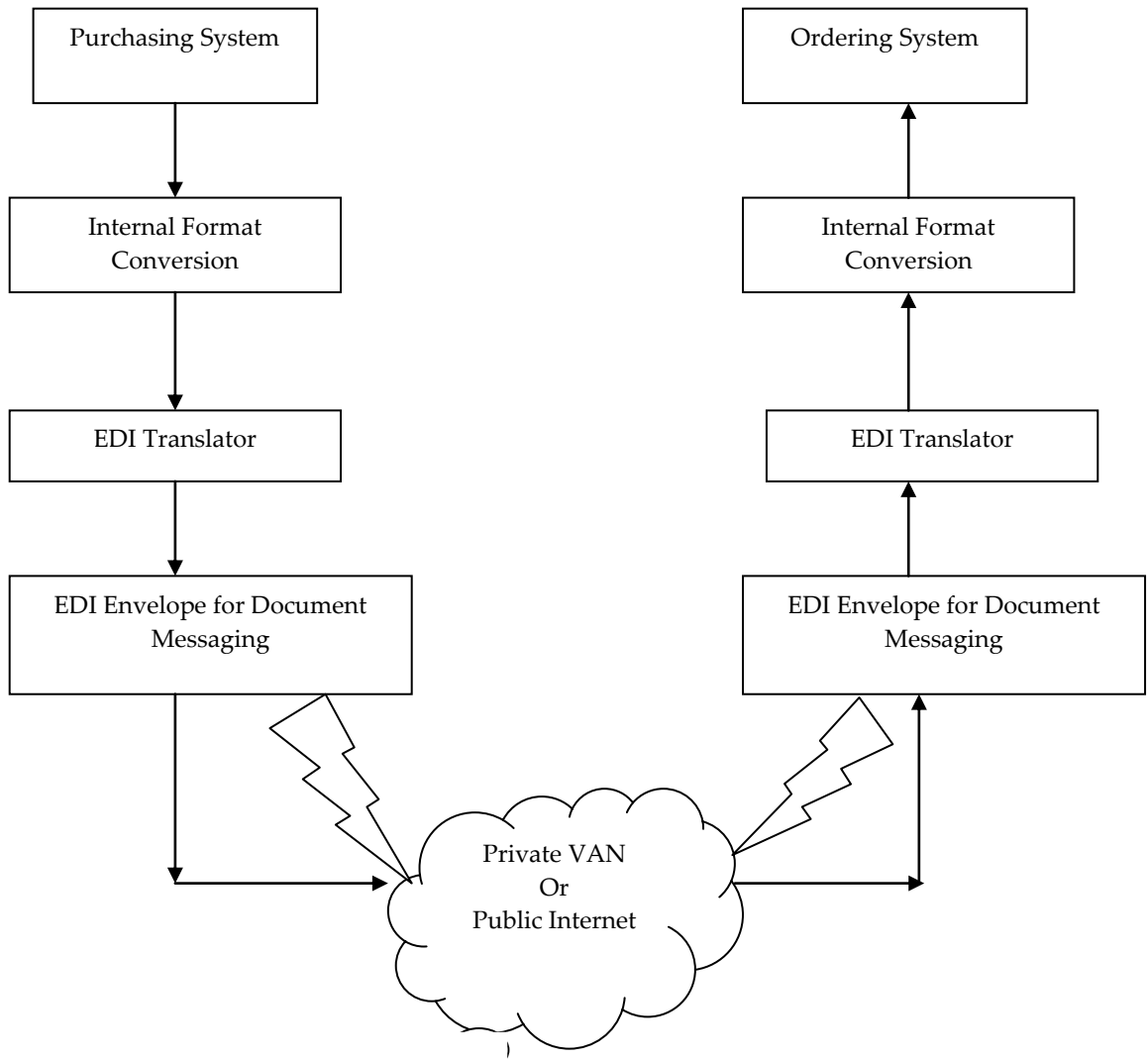

**Figure 5:** How EDI works

## 3. Program Implementation and Testing

Having completed the analysis and design, the developer needs to write the codes that will make use of the design specifications. This is known as implementation. However, certain standards and procedures must be met prior to program coding. Standards and procedures helps in proper documentation of program for easy understanding by prospective users. They also aid in translating design to codes. Standardized documentation also helps in locating errors and in making changes.

By structuring codes according to standards, the correspondence between design modules and code modules is maintained. Changes in the design therefore pose no problem to implement in the code.

In this section, we present the requirements for this application to be deployed successfully and the screen shots of some of the pages.

### 3.1  Systems requirement

### 3.1.1 Hardware requirements

a.  Central Processing Unit (CPU):
   1GHz - 2 GHz processor
b.  Memory: 1GB - 2GB of RAM
c.  Hard Disk Space: 40GB - 80GB

### 3.1.2  Software Requirement

a.  Operating System Requirements:
   Windows 2000 Professional Edition and above
   .NET Frame work: Microsoft .NET Framework 3.5
b.  Database Server: Microsoft SQL Server 2005 or 2008
c.  Web Server: Internet Information Services (IIS) Manage

### 3.2    System Implementation and Screen Shots

The system requires that both the user and the administrator login before they could be allow access. It also prompt new user to register before access is granted by the system. Once successful login by the user, the user can transfer files, view messages and update his/her profile. In addition, an administrator can also do likewise.

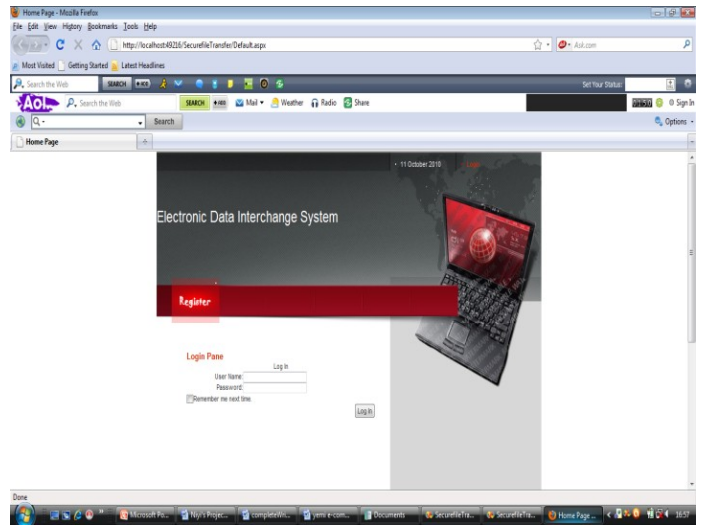

Figure 6: Login page

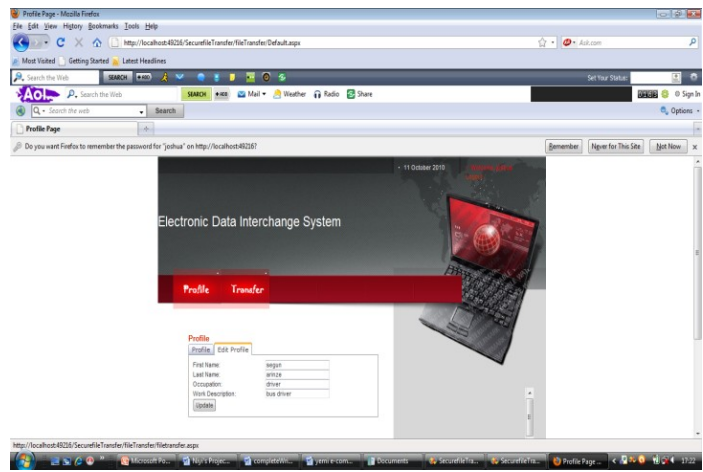

Figure 7: User Home Page

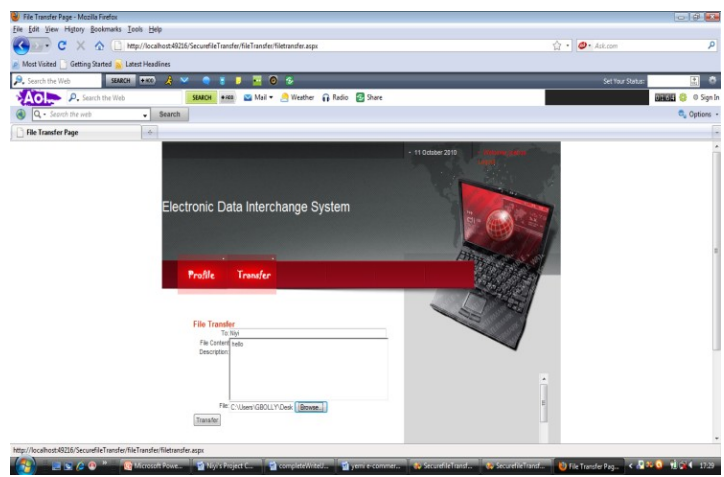

Figure 8: Sending a file

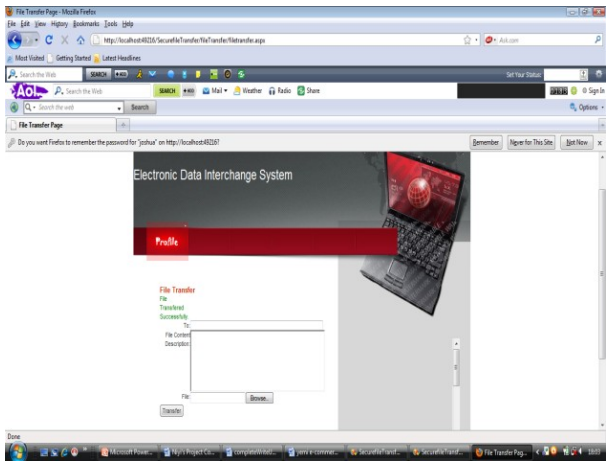

Figure 9:- File transferred successfully

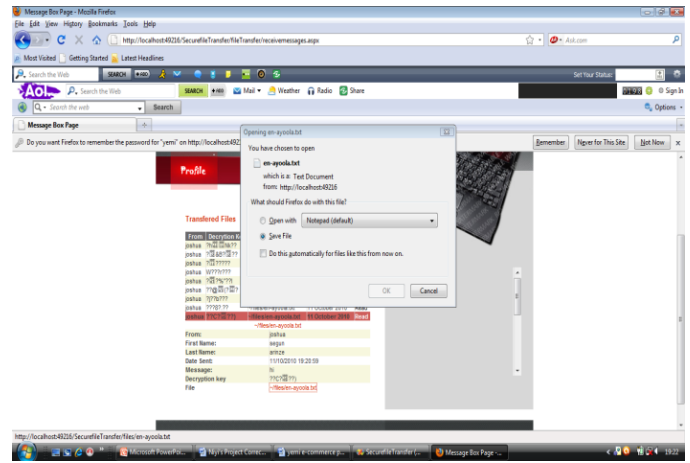

Figure 12: File download

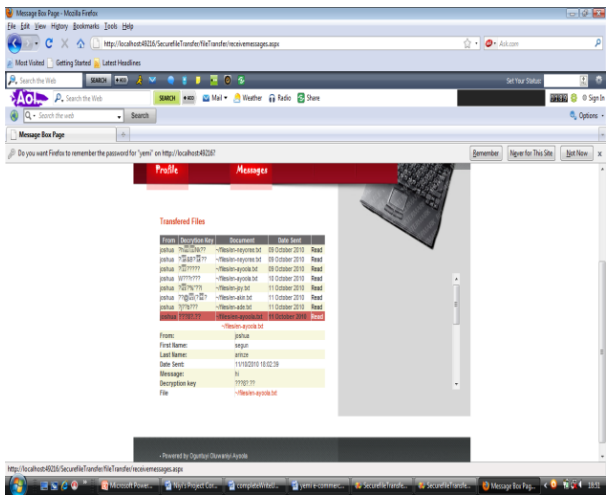

Figure 10: Transferred files

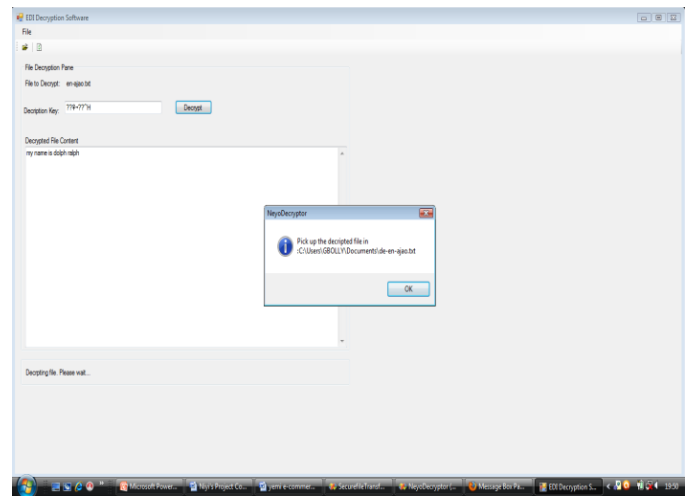

Figure 13: decrypted file

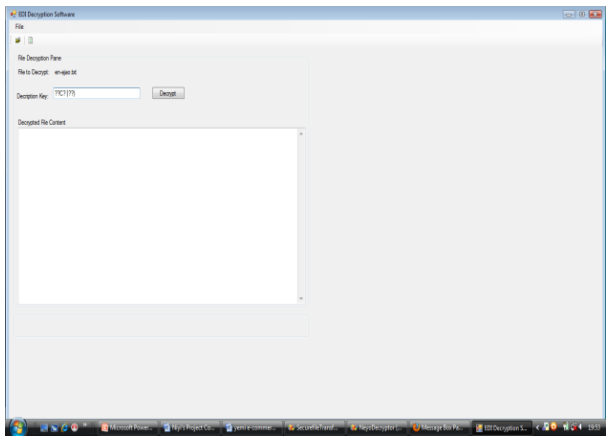

Figure 11: File to decrypt

## 4.     Conclusion

As the dramatic growth of electronic commerce on the internet progresses, the concept of using the internet for EDI has become viable idea in the industry. This paper implements a data encryption technique as an electronic data interchange security measure in the communication between computer systems.

**ACHIMUGU Philip** is a Lecturer at the Department of Computer Science of Lead City University, Ibadan, Nigeria. He holds B.Sc and M.Sc degrees in 2004 and 2009 respectively in Computer Science. He is also a PhD student at the Obafemi Awolowo University, Ile-Ife, Nigeria in the same field. He has over 18 publications in reputable journals and referred learned conferences both home and abroad. His research area is mainly software engineering with emphasis on: Development techniques, Development tools, Software products architecture and Usability.

**OLUWAGBEMI Oluwatolani** is a Lecturer at Lead City University, Ibadan, Nigeria. She can be reached at Department of Computer Science (Room 118), Faculty of Information Technology and Applied Science Building. She Holds B.Sc and M.Sc degrees in 2005 and 2011 respectively. Her research interests include computer-based information systems. She have published about 15 articles in locally and internationally reputable journals and learned conferences.

**Abah Joshua** is a Lecturer at the Department of Computer Engineering of the University of Maiduguri, Nigeria. He Holds B.Sc and M.Sc degrees in Computer Science in 2005 and 2011 respectively. His research interests include computer networks and communication. He has published many articles in locally and internationally recognized journals and learned conferences.